\documentclass[10pt]{article}
\usepackage{graphicx}
\usepackage[utf8]{inputenc}

\def\Title#1{\begin{center} {\Large #1 } \end{center}}
\def\Author#1{\begin{center}{ \sc #1} \end{center}}
\def\Address#1{\begin{center}{ \it #1} \end{center}}

\newcommand\pubblock{\rightline{\begin{tabular}{l} Proceedings of the Fifth Annual LHCP\\ \pubnumber\\
         \pubdate  \end{tabular}}}

\newenvironment{Abstract}{\begin{quotation} \begin{center} 
             \large ABSTRACT \end{center}\bigskip 
      \begin{center}\begin{large}}{\end{large}\end{center} \end{quotation}}

\newenvironment{Presented}{\begin{quotation} \begin{center} 
             PRESENTED AT\end{center}\bigskip 
      \begin{center}\begin{large}}{\end{large}\end{center} \end{quotation}}

\def\Acknowledgements{\bigskip  \bigskip \begin{center} \begin{large}
             \bf ACKNOWLEDGEMENTS \end{large}\end{center}}
\def\beq{\begin{equation}}
\def\eeq#1{\label{#1}\end{equation}}
\def\eeqn{\end{equation}}

\def\beqa{\begin{eqnarray}}
\def\eeqa#1{\label{#1}\end{eqnarray}}
\def\eeqan{\end{eqnarray}}

\let\bar=\overbar

\def\Dslash{\not{\hbox{\kern-4pt $D$}}}
\def\dslash{\not{\hbox{\kern-2pt $\del$}}}

\def\msb{{\bar{\ssstyle M \kern -1pt S}}}

\usepackage{color}
\usepackage{amsmath}
\newcommand\pubnumber{ CMS CR-2017/305 }
\newcommand\pubdate{\today}

\def\affiliation{
On behalf of the CMS collaboration, \\
State Key Laboratory of Nuclear Physics and Technology, Peking University \\
100871 Beijing, China}

\begin{document}

% large size for the first page
\large
\begin{titlepage}
\pubblock

%% Change the title, name, abstract
%% Title 
\vfill
\Title{ Rare B Decays at CMS }
\vfill

%  if you need to add the support use this, fill the \support definition above. 
%   \Author{ FIRSTNAME LASTNAME \support }
\Author{ LINWEI LI  }
\Address{\affiliation}
\vfill
\begin{Abstract}

The flavour changing neutral current decays can be interesting probes for searching for New Physics. Angular distributions of the decay $\mathrm{B}^0 \to \mathrm{K}^{*0} \mu^ +\mu^-$ are studied using a sample of proton-proton collisions at $\sqrt{s} = 8~\mathrm{TeV}$ collected with the CMS detector at the LHC, corresponding to an integrated luminosity of $20.5~\mathrm{fb}^{-1}$. An angular analysis is performed to determine $P_1$ and $P_5'$, where $P_5'$ is of particular interest due to recent measurements that indicate a potential discrepancy with the standard model. Based on a sample of 1397 signal events, $P_1$ and $P_5'$ angular parameters are determined as a function of the dimuon invariant mass squared. The measurements are in agreement with standard model predictions.

\end{Abstract}
\vfill

% DO NOT CHANGE 
\begin{Presented}
The Fifth Annual Conference\\
 on Large Hadron Collider Physics \\
Shanghai Jiao Tong University, Shanghai, China\\ 
May 15-20, 2017
\end{Presented}
\vfill
\end{titlepage}
\def\thefootnote{\fnsymbol{footnote}}
\setcounter{footnote}{0}
%

% normal size for the rest
\normalsize 

%% Your paper should be entered below. 

\section{Introduction}

Phenomena beyond the standard model (SM) of particle physics can become manifest directly, via the production of new particles, or indirectly, by affecting the production and decay of SM particles. Rare flavour changing neutral current (FCNC) decays constitute sensitive probes for New Physics (NP) since they are forbidden at tree-level in the Standard Model (SM) but can be described by box or penguin diagrams. An example is the decay $\mathrm{B}^0 \to \mathrm{K}^{*0} \mu^ +\mu^-$, where $\mathrm{K}^{*0}$ indicates the $\mathrm{K}^{*0}(892)$ meson, which provides many opportunities to search for new phenomena.

The differential decay rate for $\mathrm{B}^0 \to \mathrm{K}^{*0} \mu^ +\mu^-$ can be written in terms of $q^2$ and three angular variables as a combination of spherical harmonics. New physics may modify any of the angular variables~\cite{
  DescotesGenon:2012zf} relative to their SM values\cite{Jager:2012uw, Descotes-Genon:2013vna}.
While previous measurements of some of these quantities by the BaBar, Belle, CDF, LHCb, and CMS experiments are
consistent with the SM\cite{Khachatryan:2015isa}, the LHCb and Belle Collaborations recently reported a discrepancy larger than 3 standard deviations $(\sigma)$ with respect to the
SM for the so-called $P_5'$ variable~\cite{Aaij:2015oid,Wehle:2016yoi}.

The new measurements of the $P_1$ and $P_5'$ angular parameters in the decay $\mathrm{B}^0 \to \mathrm{K}^{*0} \mu^ +\mu^-$~\cite{Sirunyan:2017dhj} are presented, using a sample of events collected in proton-proton (pp) collisions at a center-of-mass energy of $8~\mathrm{TeV}$ with the CMS detector at LHC. The data correspond to an integrated luminosity of $20.5~\mathrm{fb}^{-1}$~\cite{CMS:2013gfa}. The value of $P_1$ and $P_5'$ angular parameters are measured by fitting the distribution of events as a function of the three angular variables that are explained in Sec.~\ref{sec:formular}. All measurements are performed in $q^2$ bins from 1 to $19~{GeV}^2$. The $q^2$ bins
$8.68<q^2<10.09~{GeV}^2$ and $12.90<q^2<14.18~{GeV}^2$, corresponding to $\mathrm{B}^0 \to \mathrm{K}^{*0} J/\psi$ and $\mathrm{B}^0 \to \mathrm{K}^{*0} \mathrm{\psi}^{'}$ decays, respectively, are used to validate the analysis.

\section{Analysis method}
\label{sec:formular}

\begin{figure}[htb]
\centering
\includegraphics[width=0.99\textwidth]{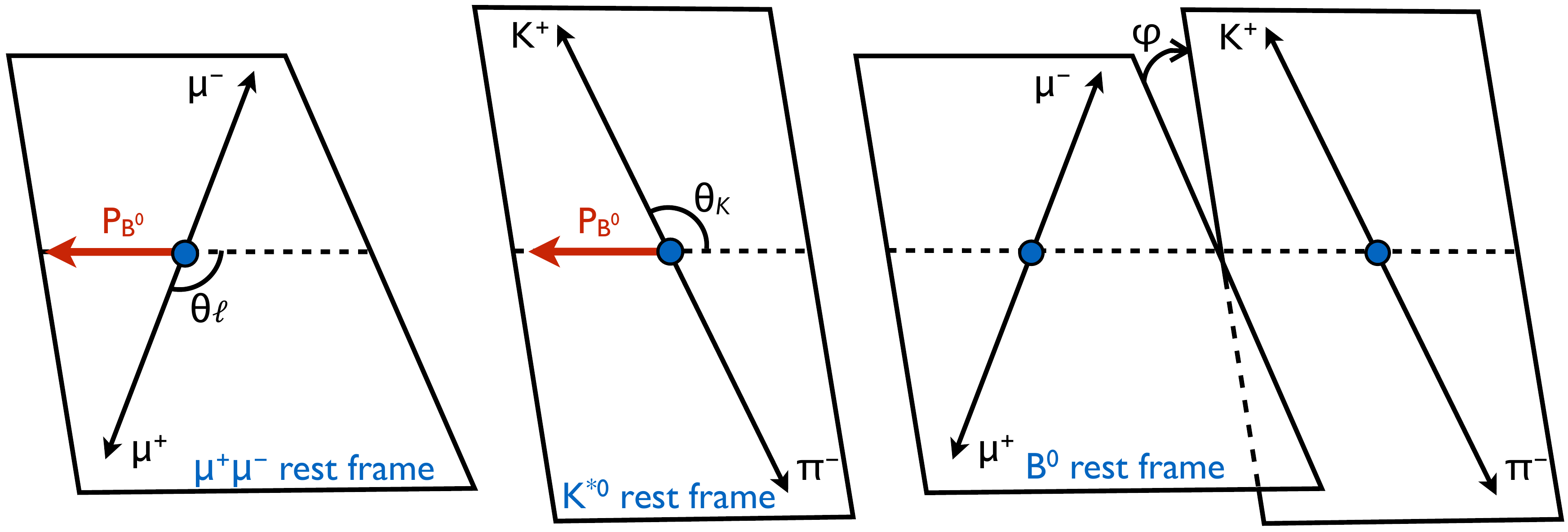}
\caption{ Sketch showing the definition of the angular variables $\theta_\ell$ (left), $\theta_K$ (middle), and $\varphi$ (right) for the decay $\mathrm{B}^0 \to \mathrm{K}^{*0} \mu^ +\mu^-$.}
\label{fig:ske}
\end{figure}

This analysis measures the $P_1$ and $P_5'$ variables in $\mathrm{B}^0 \to \mathrm{K}^{*0} \mu^ +\mu^-$ decay as a function of $q^2$.  Figure~\ref{fig:ske} shows the angular variables needed to
define the decay: $\theta_\ell$ is the angle between the positive (negative) muon momentum and the direction opposite to the $\mathrm{B}^0$ ($\bar{\mathrm{B}^0}$) in the dimuon rest frame, $\theta_K$ is the angle between the kaon momentum and the direction opposite to the $\mathrm{B}^0$ ($\bar{\mathrm{B}^0}$) in the $\mathrm{K}^{*0}$ ($\bar{\mathrm{K}^{*0}}$) rest frame, and $\varphi$ is the angle between the plane containing the two muons and the plane containing the
kaon and pion in the $\mathrm{B}^0$ rest frame.

  \subsection{The probability density function}
Although the $\mathrm{K}^{+}\mathrm{\pi}^{-}$ invariant mass must be consistent with that of a $\mathrm{K}^{*0}$, there can be a contribution from spinless (S-wave) $\mathrm{K}^{-}\mathrm{\pi}^{+}$ combinations~\cite{Descotes-Genon:2013vna}. This is parametrized with three terms: $F_\mathrm{S}$, which is related to the S-wave fraction, and $A_\mathrm{S}$ and $A^5_\mathrm{S}$, which are the interference amplitudes between the S-wave and
P-wave decays. Including these components, the angular distribution of $\mathrm{B}^0 \to \mathrm{K}^{*0} \mu^ +\mu^-$ can be written as~\cite{Descotes-Genon:2013vna}:

\begin{equation} \label{eq:PDF}
 \begin{split}
   \frac{1} {\mathrm{d} \Gamma / \mathrm{d}q^2} \frac{\mathrm{d}^4\Gamma} {\mathrm{d}q^2 \mathrm{d}\cos\theta_\ell \mathrm{d}\cos\theta_\mathrm{K} \mathrm{d}\varphi} =
 & \frac{9} {8\pi} \left\{ \frac{2}{3} \left[ (F_\mathrm{S}+A_\mathrm{S}\cos\theta_\mathrm{K}) \left( 1-\cos^2\theta_\ell \right) + A^5_\mathrm{S} \sqrt{1-\cos^2\theta_\mathrm{K}} \right. \right. \\
 & \left. \sqrt{1-\cos^2\theta_\ell}\cos\varphi \right] + \left(1 - F_\mathrm{S} \right) \Bigl[ 2F_\mathrm{L}\cos^2\theta_\mathrm{K} \left( 1-\cos^2\theta_\ell \right) \Bigr. \\
 & + \frac{1} {2} \left( 1-F_\mathrm{L} \right) \left( 1-\cos^2\theta_\mathrm{K} \right) \left( 1+\cos^2\theta_\ell \right) + \frac{1} {2} P_1(1-F_\mathrm{L}) \\
 & (1-\cos^2\theta_\mathrm{K})(1-\cos^2\theta_\ell)\cos2 \varphi + 2P_5'\cos\theta_\mathrm{K} \sqrt{F_\mathrm{L} \left( 1-F_\mathrm{L} \right) } \\
 & \Bigl. \left. \sqrt{1-\cos^2\theta_\mathrm{K}} \sqrt{1-\cos^2\theta_\ell}\cos\varphi \Bigr] \right\}.
  \end{split}
\end{equation}
where $F_\mathrm{L}$ denotes the longitudinal polarization fraction of the $\mathrm{K}^{*0}$. This expression is an exact simplification of the full angular distribution, obtained by folding the $\varphi$ and
$\theta_\ell$ angles about zero and $\pi/2$, respectively. Specifically, if $\varphi < 0$, then $\varphi \to -\varphi$, and the new $\varphi$ domain is [0, $\pi$]. If $\theta_\ell > \pi/2$, then $\theta_\ell \to \pi - \theta_\ell$, and the new $\theta_\ell$ domain is [0, $\pi/2$].
We use this simplified version of the expression because of difficulties in the fit convergence with the full angular distribution due to the limited size of the data sample. This simplification exploits the odd symmetry of the angular variables with respect
to $\varphi = 0$ and $\theta_\ell = \pi/2$ in such a manner that the cancellation around these angular values is exact. This cancellation remains approximately valid even after accounting for the experimental acceptance because the efficiency is symmetric
with respect to the folding angles.

For each $q^2$ bin, the observables of interest are extracted from an unbinned extended maximum-likelihood fit to four variables: the $\mathrm{K}^{+}\mathrm{\pi}^{-} \mu^ +\mu^-$ invariant mass $m$ and the three angular variables ${\theta_\ell}$,
${\theta_K}$, and $\varphi$. For each $q^2$ bin, the unnormalized probability density function (pdf) has the following expression:

\begin{equation} \label{eq:angALL}
  \begin{split}
    \mathrm{pdf}(m,\theta_K,\theta_\ell,\varphi) & = Y^{C}_{S} \biggl[ S^{C}(m)  \, S^a(\theta_K,\theta_\ell,\varphi) \, \epsilon^{C}(\theta_K,\theta_\ell,\varphi) \biggr. \\
    & \biggl. + \frac{f^{M}}{1-f^{M}}~S^{M}(m) \, S^a(-\theta_K,-\theta_\ell,\varphi) \, \epsilon^{M}(\theta_K,\theta_\ell,\varphi) \biggr] \\
    & + Y_{B}\,B^m(m) \, B^{\theta_K}(\theta_K) \, B^{\theta_\ell}(\theta_\ell) \, B^{\varphi}(\varphi), \\
  \end{split}
\end{equation}

where the contributions correspond to correctly tagged signal events, mistagged signal events, and background events. The parameters $Y^{C}_{S}$ and $Y_{B}$ are the yields of correctly tagged signal
events and background events, respectively, and are free parameters in the fit. The parameter $f^{M}$ is the fraction of signal events that are mistagged and is determined from MC simulation. The signal mass probability functions $S^{C}(m)$ and $S^{M}(m)$ are each the sum of two Gaussian functions sharing the same mean, and describe the mass distribution for correctly tagged and mistagged signal events, respectively.

In the fit, the mean, the four Gaussian $\sigma$ parameters, and two fractions relating the contribution of each Gaussian, are determined from simulation, which has been found to accurately reproduce the data. The function $S^a(\theta_K,\theta_\ell,\varphi)$ describes the signal in the three-dimensional (3D) space of the angular variables and corresponds to Eq.~(\ref{eq:PDF}).
The combination $B^m(m) \, B^{\theta_K}(\theta_K) \, B^{\theta_\ell}(\theta_\ell) \, B^{\varphi}(\varphi)$ is obtained from $\mathrm{B}^0$ sideband data and describes the background in the space of $(m,\theta_K,\theta_\ell,\varphi)$,
where the mass distribution is an exponential function and the angular distributions are polynomials ranging from second to fourth degree, for both $B^{\theta_K}(\theta_K)$ and $B^{\theta_\ell}(\theta_\ell)$, depending on the $q^2$
bin, while the term $B^{\varphi}(\varphi)$ is of first degree for all $q^2$ bins.

The functions $\epsilon^{C}(\theta_K,\theta_\ell,\varphi)$ and $\epsilon^{M}(\theta_K,\theta_\ell,\varphi)$ are the efficiencies in the 3D space of $-1 \leq \cos\theta_K \leq 1, 0 \leq \cos\theta_\ell\leq 1$, and $0\leq \varphi\leq \pi$ for correctly tagged and mistagged signal events, respectively. For each $q^2$, we compute both correctly tagged events efficiency and mistagged events efficiency. The numerator and denominator of the efficiency are separately described with a nonparametric technique, which is implemented with a kernel density estimator. The final efficiency distributions used in the fit are obtained from the ratio of 3D histograms derived from the sampling of the kernel density estimators. The histograms have 40 bins in each dimension. A consistency check of the procedure used to determine the efficiency is performed by dividing the simulated data sample into two independent subsets, and extracting the angular variables from the first subset using the efficiency computed from the second one. 

 \subsection{The fitting sequence}
 
The fit is performed in two steps. 
The initial fit uses the data from the sidebands of the $\mathrm{B}^0$ mass to obtain the $B^m(m)$, $B^{\theta_K}(\theta_K)$, $B^{\theta_\ell}(\theta_\ell)$, and $B^{\varphi}(\varphi)$ distributions (the signal component is absent from this fit). The sideband regions are $3\sigma_{m} < |m-m_{\mathrm{B}^0}| < 5.5\sigma_{m}$, where $\sigma_m$ is the average mass resolution ($\approx$45MeV), obtained from fitting the MC simulation signal to a sum of two Gaussians with a common mean. The distributions obtained in this step are then fixed for the second step, which is a fit to the data over the full mass range. The free parameters in this fit are the angular parameters $P_1$, $P_5'$, and $A^5_\mathrm{S}$, and the yields $Y^{C}_{S}$ and $Y_{B}$.

In order to avoid fit convergence problems due to the limited number of signal candidate events, the angular parameters $F_\mathrm{L}$, $F_\mathrm{S}$, and $A_\mathrm{S}$ are fixed to previous CMS measurements performed
on the same data set with the same event selection criteria~\cite{Khachatryan:2015isa}.

To ensure correct coverage for the uncertainties of the angular parameters, the Feldman-Cousins (FC) method~\cite{Feldman:1997qc} is used with nuisance parameters.
Two main sets of pseudo-experimental samples are generated to compute the coverage for the two angular observables $P_1$ and $P_5'$, respectively. The first (second) set, used to compute the coverage for $P_1$ ($P_5'$), is generated
by assigning values to the other parameters as obtained by profiling the bivariate Normal distribution description of the likelihood on data at fixed $P_1$ ($P_5'$) values. When fitting the pseudo-experimental samples, the same fit procedure
as applied to data is used.

The fit formalism and results are validated through fits to pseudo-experimental samples, MC simulation samples, and control channels.

\section{Systematic uncertainties}
\label{sec:Systematics}

Since the efficiency is computed with simulated events, it is essential that the MC simulation program correctly reproduce the data, and extensive checks have been performed to verify its accuracy. The systematic uncertainties are described below and summarized in Table~\ref{tab:systematics}.

The possible sources of systematic uncertainties investigated are:
\begin{description}
    \item[simulation mismodeling:] the results from the fitting on the
      generated pure signal events are used to estimate simulation mis-modeling;

    \item[fit bias:] the possible biases from the fitting programs
    and procedures;
    
    \item[MC statistical uncertainty:] the efficiency is computed using
      a finite set of simulated events. The size of this
      sample affects the accuracy of the determination of the efficiency;

    \item[efficiency:] the efficiency shape is validated using
      the control samples;

    \item[wrong $K\pi$ assignment:] the effect of wrong $K\pi$ assignment on the event
      interpretation and fitting results;

    \item[background distributions:] the signal region has a background
      contamination. The description of the background distributions
      will affect the fitting results;

    \item[mass distributions:] the parameters of the mass distribution
     are fixed in the final fitting, and their uncertainties
      need to be propagated to the results;

   \item [feed-through background:] the $q^2$ bins just below and above the
      resonance regions of $J/\psi$ and $\psi'$ may be
      contaminated with $\mathrm{B}^0 \to \mathrm{K}^{*0} J/\psi$ and $\mathrm{B}^0 \to \mathrm{K}^{*0} \mathrm{\psi}^{'}$
      feed-through events that are not removed by the selection criteria;
      
\item [$F_\mathrm{L}$, $F_\mathrm{S}$, $A_\mathrm{S}$ uncertainty propagation:] in the final fit, we fix the parameters, $F_\mathrm{L}$, $F_\mathrm{S}$, $A_\mathrm{S}$, at the previous CMS measurements~\cite{Khachatryan:2015isa}. Their uncertainties are propagated to the final results;
   
 \item[angular resolution:] the fitting is performed on reconstructed
      quantities, whose resolution does affect the fitting results.

\end{description}

\begin{table*}[t]
  \begin{center}
  \caption {Systematic uncertainty contributions for the measurements of
    $P_1$ and $P_5'$.
    The total uncertainty in each $q^2$ bin is obtained by adding each contribution in quadrature.
    For each item, the range indicates the variation of the uncertainty in the $q^2$ bins.}
\begin{tabular}{l|cccc}
Systematic uncertainty & $P_1 (10^{-3})$ & $P_5' (10^{-3})$ \\[1pt]
\hline \\[-2ex]
Simulation mismodeling       &   1--33   &  10--23  \\[1pt]
Fit bias                     &   5--78   &  10--119 \\[1pt]
MC statistical uncertainty   &  29--73   &  31--112 \\[1pt]
Efficiency                   &  17--100  &   5--65  \\[1pt]
$K\pi$ mistagging          &   8--110  &   6--66  \\[1pt]
Background distribution      &  12--70   &  10--51  \\[1pt]
Mass distribution            &      12   &      19  \\[1pt]
Feed-through background      &   4--12   &   3--24  \\[1pt]
$F_\mathrm{L}$, $F_\mathrm{S}$, $A_\mathrm{S}$ uncertainty propagation & 0--126 & 0--200 \\[1pt]
Angular resolution           &   2--68   & 0.1--12  \\[1pt]
\hline
Total systematic uncertainty &  60--220  &  70--230 \\[1pt]
\end{tabular}
\label{tab:systematics}
\end{center}
\end{table*}

\section{Fitting result}

\begin{figure*}[htbp]
  \begin{center}
    \includegraphics[width=0.8\textwidth]{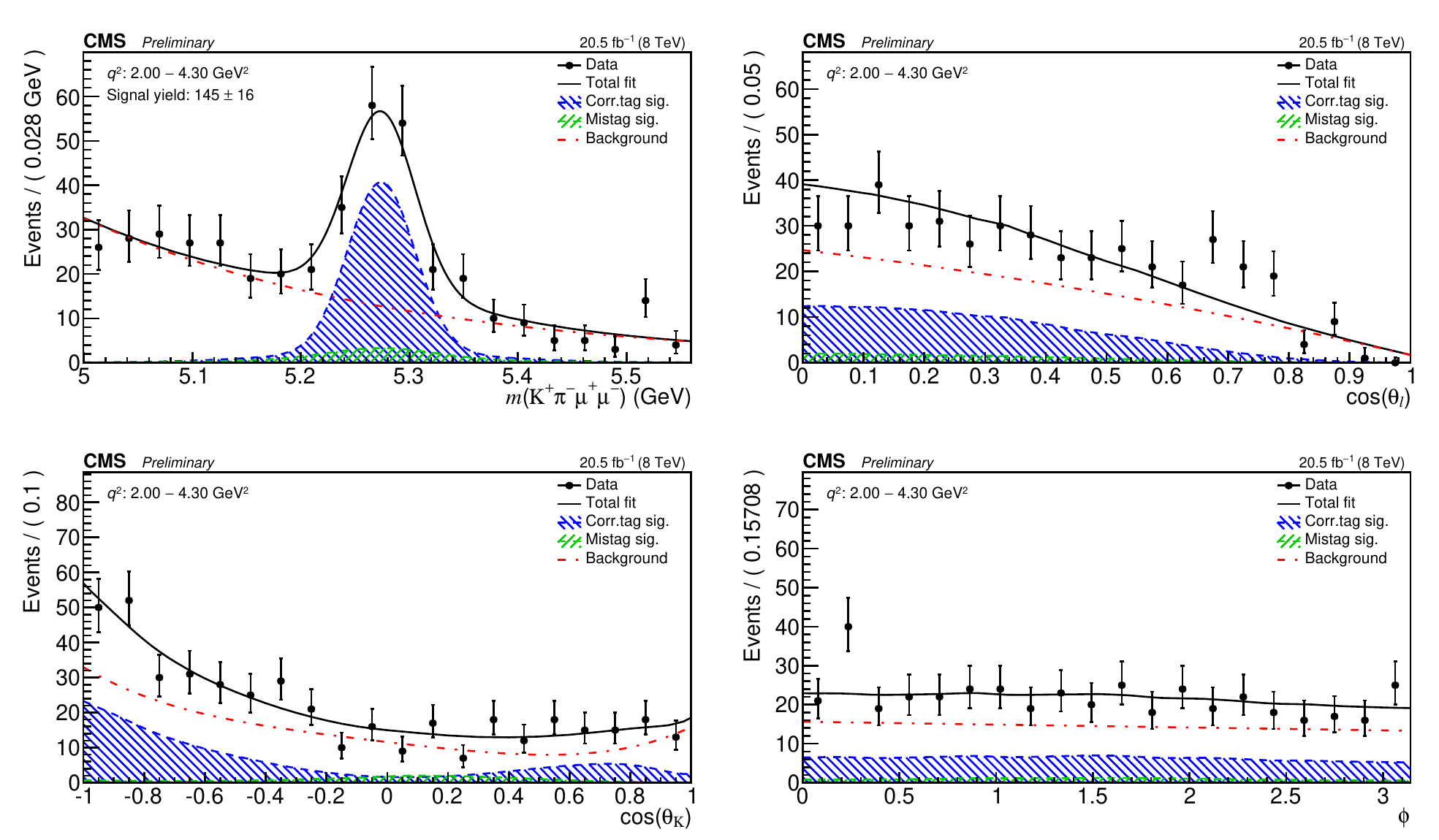}
    \caption{$\mathrm{K}^{+}\mathrm{\pi}^{-} \mu^ +\mu^-$ invariant mass and angular distributions for the second $q^2$ bin
      $2.00<q^2<4.30~{GeV}^2$.
      Overlaid on each plot is the projection of the results for the total fit, as well as for the three components: correctly tagged
      signal, mistagged signal, and background. The vertical bars indicate the statistical uncertainties~\cite{Sirunyan:2017dhj}.}
    \label{fig:fullplots}
  \end{center}
\end{figure*}

\begin{figure}[htbp!]
  \begin{center}
    \includegraphics[width=0.49\textwidth]{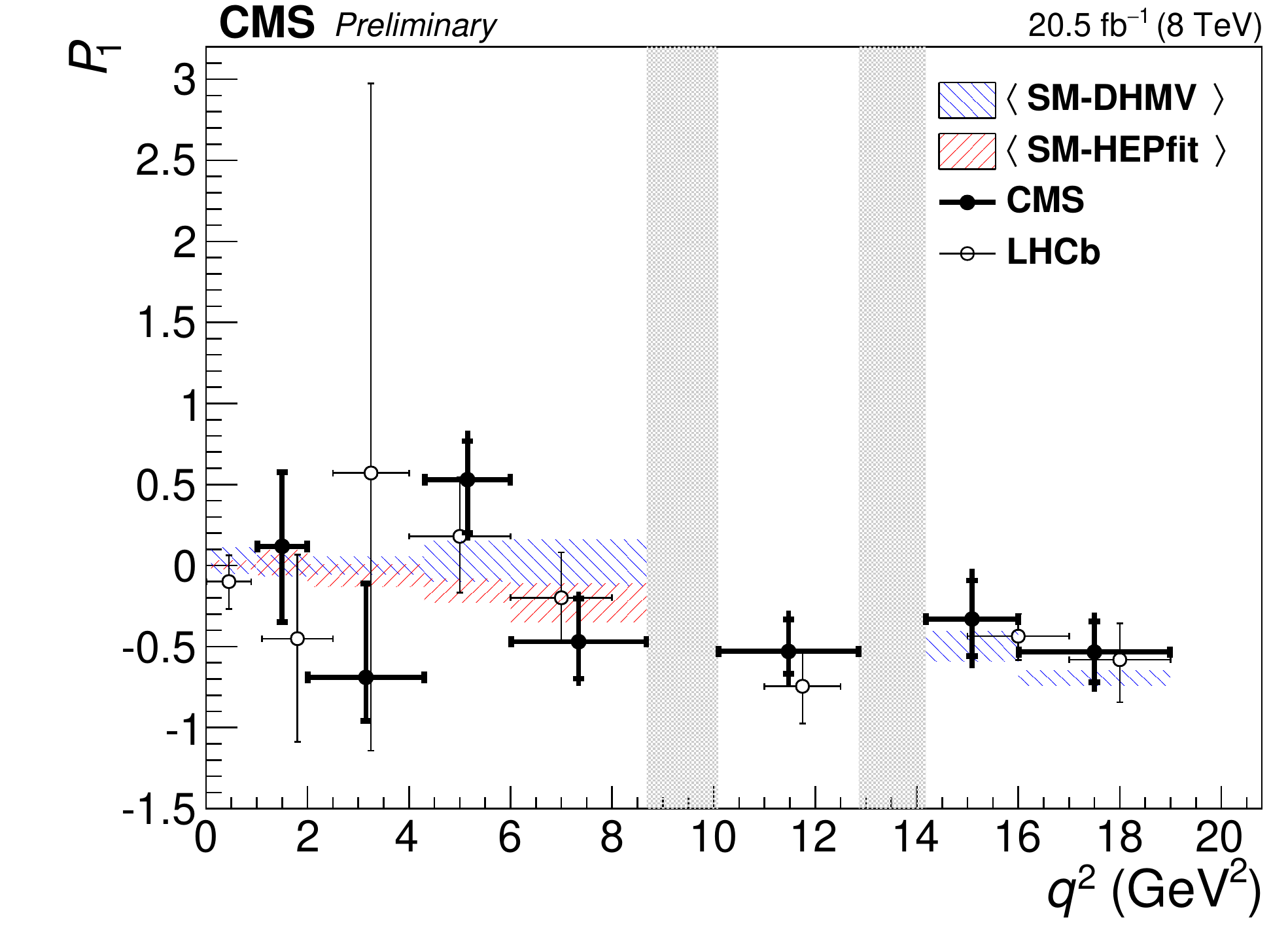}
    \includegraphics[width=0.49\textwidth]{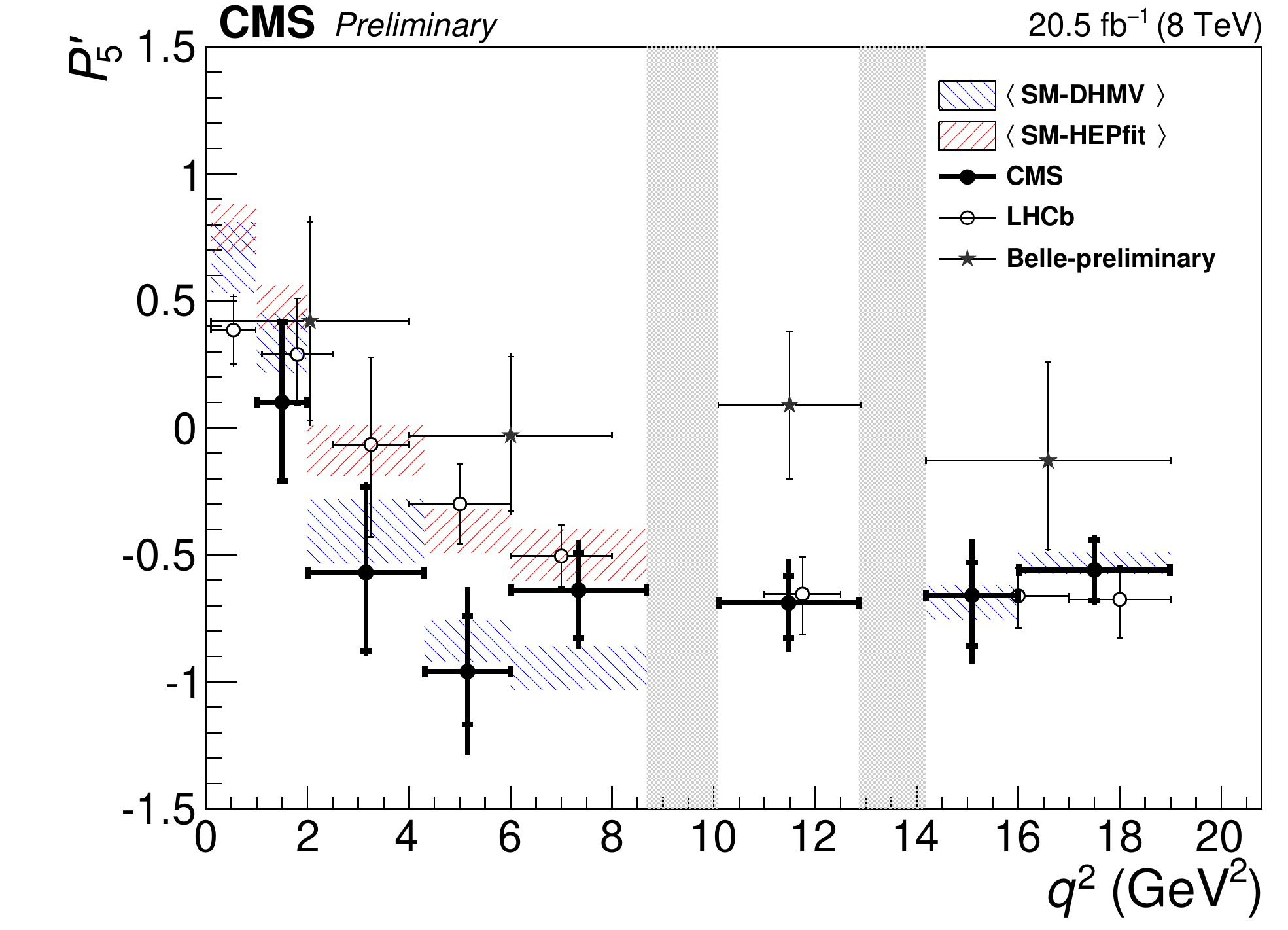}
    \caption{Measured values of $P_1$ and $P_5'$ versus $q^2$ for $\mathrm{B}^0 \to \mathrm{K}^{*0} \mu^ +\mu^-$ from CMS~\cite{Sirunyan:2017dhj}, compared with LHCb~\cite{Aaij:2015oid}
      and Belle~\cite{Wehle:2016yoi} results. The statistical uncertainty is shown by the inner vertical bars, while the outer vertical bars give the total uncertainty. The horizontal bars show the bin widths. The vertical shaded regions correspond to the $J/\psi$ and $\psi'$ resonances. The red and blue hatched regions show two SM predictions averaging over each $q^2$ bin to provide a direct comparison to the data. The SM-DHMV result is derived from Refs.~\cite{DescotesGenon:2012zf,Descotes-Genon:2013vna}, while SM-HEPfit result from Refs.~\cite{Ciuchini:2015qxb,Ciuchini:2016weo}. Reliable theoretical predictions are not available near the $J/\psi$  and $\psi'$ resonances.}
    \label{fig:results}
  \end{center}
\end{figure}

The signal data, corresponding to 1397 events, are fit in seven $q^2$ bins from 1 to $19~{GeV}^2$. As an example, distributions for the second $q^2$ bin, along with the fit projections, are shown in Fig.~\ref{fig:fullplots}. The fitted values of $P_1$, and $P_5'$, along with their associated uncertainties, for each of the $q^2$ regions are shown in Fig.~\ref{fig:results}, along with the SM predictions. 

%\begin{table*}[htb]
  %\centering
%  \caption{The measured signal yields, which include both correctly tagged and mistagged events,
 %   and the $P_1$ and $P_5'$ values, in bins of $q^2$, for the decay $\mathrm{B}^0 \to \mathrm{K}^{*0} \mu^ +\mu^-$. The first uncertainty is statistical
 %   and the second is systematic. The bin ranges are  selected to allow comparisons to previous measurements.}
%\begin{tabular}{c|ccc}
%$q^2~({GeV}^2)$& Signal yield& $P_1$                            & $P_5'$ \\[1pt]
%\hline \\[-2ex]
%1.00--2.00     & $ 80 \pm 12$ & $+0.12^{+0.46}_{-0.47} \pm 0.06$ & $+0.10^{+0.32}_{-0.31} \pm 0.12$ \\[1pt]
%2.00--4.30     & $145 \pm 16$ & $-0.69^{+0.58}_{-0.27} \pm 0.09$ & $-0.57^{+0.34}_{-0.31} \pm 0.15$ \\[1pt]
%4.30--6.00     & $119 \pm 14$ & $+0.53^{+0.24}_{-0.33} \pm 0.18$ & $-0.96^{+0.22}_{-0.21} \pm 0.16$ \\[1pt]
%6.00--8.68     & $247 \pm 21$ & $-0.47^{+0.27}_{-0.23} \pm 0.13$ & $-0.64^{+0.15}_{-0.19} \pm 0.14$ \\[1pt]
%10.09--12.86   & $354 \pm 23$ & $-0.53^{+0.20}_{-0.14} \pm 0.14$ & $-0.69^{+0.11}_{-0.14} \pm 0.23$ \\[1pt]
%14.18--16.00   & $213 \pm 17$ & $-0.33^{+0.24}_{-0.23} \pm 0.22$ & $-0.66^{+0.13}_{-0.20} \pm 0.19$ \\[1pt]
%16.00--19.00   & $239 \pm 19$ & $-0.53^{+0.19}_{-0.19} \pm 0.13$ & $-0.56^{+0.12}_{-0.12} \pm 0.07$ \\[1pt]
%\hline
%\end{tabular}
%\label{tab:results} 
%\end{table*}

Two SM predictions, SM-DHMV and SM-HEPfit, are available for comparison with the measured angular parameters. The SM-DHMV result, derived from Refs.~\cite{DescotesGenon:2012zf,Descotes-Genon:2013vna}, updates the calculations from Ref.~\cite{Ball:2004rg} to account for the known correlation between the different form factors.

Light-cone sum rule predictions, which are valid in the low-$q^2$ region, are also combined with lattice determinations at high $q^2$ to yield more precise determinations of the form factors over the full $q^2$ range. The hadronic charm-loop contribution is derived from Ref.~\cite{Khodjamirian:2010vf}. The SM-HEPfit result, derived from the calculation reported in Refs.~\cite{Ciuchini:2015qxb,Ciuchini:2016weo}, uses full QCD form factors and derives the hadronic contribution from LHCb data~\cite{Aaij:2015oid}.

Reliable theoretical predictions are not available near the $J/\psi$ and $\psi'$ resonances. The two SM predictions are shown in comparison to the data in Fig.~\ref{fig:results}. Both are seen to be in agreement with the CMS results, although the agreement with SM-DHMV is somewhat better. Thus we do not obtain evidence for physics beyond the SM. Qualitatively, the LHCb data appear to be in better agreement with the SM-HEPfit prediction than with SM-DHMV result, but the uncertainties are too large to allow a definite conclusion.

\section{Conclusions}

Using pp collision data recorded at $\sqrt{s}=8~\mathrm{TeV}$ with the CMS detector at the LHC, corresponding to an integrated luminosity of $20.5~\mathrm{fb}^{-1} $ an angular analysis has been performed for the decay $\mathrm{B}^0 \to \mathrm{K}^{*0} \mu^ +\mu^-$ . In total, 1397 signal events are obtained. For each bin of the dimuon invariant mass squared $(q^2)$, unbinned maximum-likelihood fits were performed to the distributions of the $\mathrm{K}^{+}\mathrm{\pi}^{-} \mu^ +\mu^-$ invariant mass and the three decay angles, to obtain values of the $P_1$ and $P_5'$ parameters. The results are among the most precise to date and are consistent with standard model predictions and previous measurements.

%%  if necessary
\Acknowledgements
I would like to thank the LHCP2017 organizers for their hospitality and the wonderful working environment. I acknowledge the support from the National Natural Science Foundation of China, under Grants No.  11661141008.

\end{document}